\documentclass[aps,pra,twocolumn]{revtex4-2}
\usepackage{amsfonts}
\usepackage{amsmath}
\usepackage{mathrsfs}
\usepackage{amssymb}
\usepackage{graphicx}
\usepackage{lipsum} 
\usepackage{bm}
\usepackage{color}

\usepackage[colorlinks=true,linkcolor=blue,anchorcolor=blue,citecolor=blue,urlcolor=blue]{hyperref}

\newcommand{\bPsi}{{\bm \Psi}}
\newcommand{\bu}{{\bm u}}
\newcommand{\bvarphi}{{\bm \varphi}}

\newcommand{\rev}[1]{\textcolor{red}{{#1}}}

\begin{document}

\title{Stabilizing period-doubled density waves by spin-orbit coupling in Bose-Einstein condensates in optical lattices}

\author{Chenhui Wang}
\affiliation{Institute for Quantum Science and Technology, Department of Physics,  Shanghai University, Shanghai 200444, China}

\author{Yongping Zhang}
\email{yongping11@t.shu.edu.cn}
\affiliation{Institute for Quantum Science and Technology, Department of Physics, Shanghai University, Shanghai 200444, China}

\begin{abstract}

In atomic Bose-Einstein condensates in optical lattices, mean-field energy can support the existence of period-doubled density waves, which are similar to Bloch waves but have the double periodicity of the underlying lattice potentials. However, they are dynamically unstable. Here, we propose to use the spin-orbit coupling to stabilize the period-doubled density waves. The stabilization mechanism is revealed to relate to interaction-induced spontaneous symmetry breaking of the spin-flip parity symmetry.

\end{abstract}

\maketitle

\section{Introduction}
Atomic Bose-Einstein condensates (BECs) loaded into optical lattices (OLs) have been important quantum many-body platforms for exploring fundamental physics, emphasizing the interplay between atomic interactions and periodic potentials.~\cite{Morsch2006, Bloch2008, Wu2000,Zhang2013_1,Dasgupta2016, Smerzi2002, Louis2003, Zhang2009}.
In the mean-field frame, atomic contact interactions become the mean-field nonlinearity~\cite{Pitaevskii2016}. The combination of the mean-field energy and OLs gives rise to the nonlinear Bloch band-gap spectrum with associated nonlinear Bloch waves (NBWs) which have the same periodicity as OLs. The investigations into NBWs and nonlinear spectrum attract great research attention~\cite{Qizhong2015,Watanabe2016,He2021,Hong2023}. Elementary excitation of the ground-state NBW also features a Bloch band-gap structure, with the lowest band being relevant to sound excitation~\cite{Inguscio2009}. The sound velocity has been measured to reflect superfluid density~\cite{Chauveau2023, Spielman2023}. The mean-field energy may bring dynamical and energetic instabilities to NBWs~\cite{Wu2001,Konotop2002,Smerzi2002,Wu2003,He2021}. Both the instabilities have been experimentally observed to break down the lattice superfluidity of BECs~\cite{Burger2001,Fallani2004}. The motions induced by an accelerating force have shown the asymmetric Landau-Zener behavior between the transition from the lowest to the first nonlinear Bloch bands and the reversed transition~\cite{Arimondo2003}. When the mean-field energy dominates over OLs, a loop structure appears to adhere to a certain nonlinear Bloch band~\cite{Wu2003}. Bloch waves in the loop do not have linear analogues~\cite{Porto2016}. The motions along the looped nonlinear Bloch band result in nonlinear Landau-Zener tunneling~\cite{Wu2000}. Its transition probability does not obey the conventional Landau-Zener formulae, and the transition happens even in the adiabatic limit of the accelerating force~\cite{liujie2002}. The loop-induced nonlinear Landau-Zener tunneling has been observed and explored in experiments~\cite{Yuao, Peter2020}.  

The mean-field energy can also support period-doubled or even multiple-period density waves in BECs in OLs~\cite{Machholm2004,Seaman2005,Zhang2009,Yang2018}. Akin to NBWs, the period-doubled density waves (PDDWs) are spatially extensive and periodic but have a periodicity that is twice that of OLs. {These} patterns may appear even in a superfluid Fermi gas~\cite{Watanabe2016s}. Their existence is a complete nonlinear phenomenon. Without the mean-field energy, these states can not exist. {They} may be considered as periodic trains of solitons~\cite{Seaman2005}. The relation between them and the NBWs inside the loop structure has been discussed~\cite{Machholm2004}. The construction of {these density patterns} from two different linear Bloch waves with equal energy has been systematically studied~\cite{Yang2018}. It has been found that the PDDWs are always dynamically unstable~\cite{Machholm2004,Yang2018},  which challenges experimentally direct observations. 
How to stabilize them immediately becomes an important problem. In Ref.~\cite{Maluckov2012}, Maluckov {\it et al.} reveal that the long-range dipole-dipole interactions can be used to stabilize {them} and consider such stable patterns as a kind of supersolids since they spontaneously break the translation symmetry of OLs. So far, the long-range interactions are the only reported approach for {the} stabilization.  

In this paper, we provide an alternative approach to stabilize the PDDWs. Our approach lies in spin-orbit coupling (SOC) and, therefore,  does not need to engineer interactions.  As an intrinsic interaction between the motion and spin of a particle, SOC in solid-state materials plays an essential role in many interesting physical phenomena and applications, such as the spin Hall effect~\cite{Kato2004, Bernevig2006}. Artificial SOC has been successfully realized in BECs~\cite{Lin2011,Cheuk2012,Jingzhang2012,Khamehchi2014,Chenshuai2015,Hamner2015,PanScience2016,LiNature2017,Huang2016}, providing an experimentally accessible platform for exploring exotic superfluidity~\cite{Wang2010,Ho2011,Hu2012,Liyun2013,Zheng2013,YP2016} and elementary excitations~\cite{Khamehchi2014,Chenshuai2015}. Especially, experimentally loading the spin-orbit-coupled BECs (SOC-BECs) into OLs~\cite{Hamner2015,Bersano2019} greatly stimulates research efforts on the rich emergent physics stemming from the interplay between the SOC and lattice effects~\cite{Cai2012,Cole2012,Radic2012,Kartashov2016,Luo2021,Liang2016, Martone2016, Zhoulu2019, Luo2021, Guanqiang2021,Zezyulin2022,Yang2023}. One of the outstanding features in SOC-BECs with OLs is that the lowest nonlinear Bloch band may be flat~\cite{Zhang2013}. A family of Wannier solitons can be bifurcated from this flat band~\cite{Wang2023}. An unconventional spin dynamics can be induced by the SOC in a random OL~\cite{Sherman2015,Sherman2018}.

We study the PDDWs in SOC-BECs loaded into OLs.  Without the SOC, the PDDWs are always dynamically unstable {since they have a negative effective mass}. It is known that the state with negative effective mass is unstable in the presence of repulsive contact interactions. {Their negative effective mass} is protected by the parity symmetry. The SOC leads to the parity being joined with the spin-flip. The contact interactions can spontaneously break the joint symmetry by destroying its spin-flip part. The spontaneous symmetry breaking can change the effective mass of the PDDWs from originally negative to positive. Finally, the {states} may achieve a positive effective mass, which provides a possibility of stabilization in the presence of repulsive contact interactions. 

This paper is organized as follows. In Sec.~\ref{model}, we present the theoretical frame for the study on SOC-BECs with OLs. It includes the system of the SOC-BEC in OLs, the PDDW solutions, and the linear stability analysis by the Bogoliubov-de Gennes equation. In Sec.~\ref{state}, we reveal the spontaneous symmetry breaking of the PDDWs.  In Sec.~\ref{stability}, the stabilization mechanism is addressed. We systematically check their stability in the full parameter space.  Finally, the conclusion is provided in Sec.~\ref{conclusion}.

\section{model}
\label{model}

\begin{figure*}[t]
	\includegraphics[width=2\columnwidth]{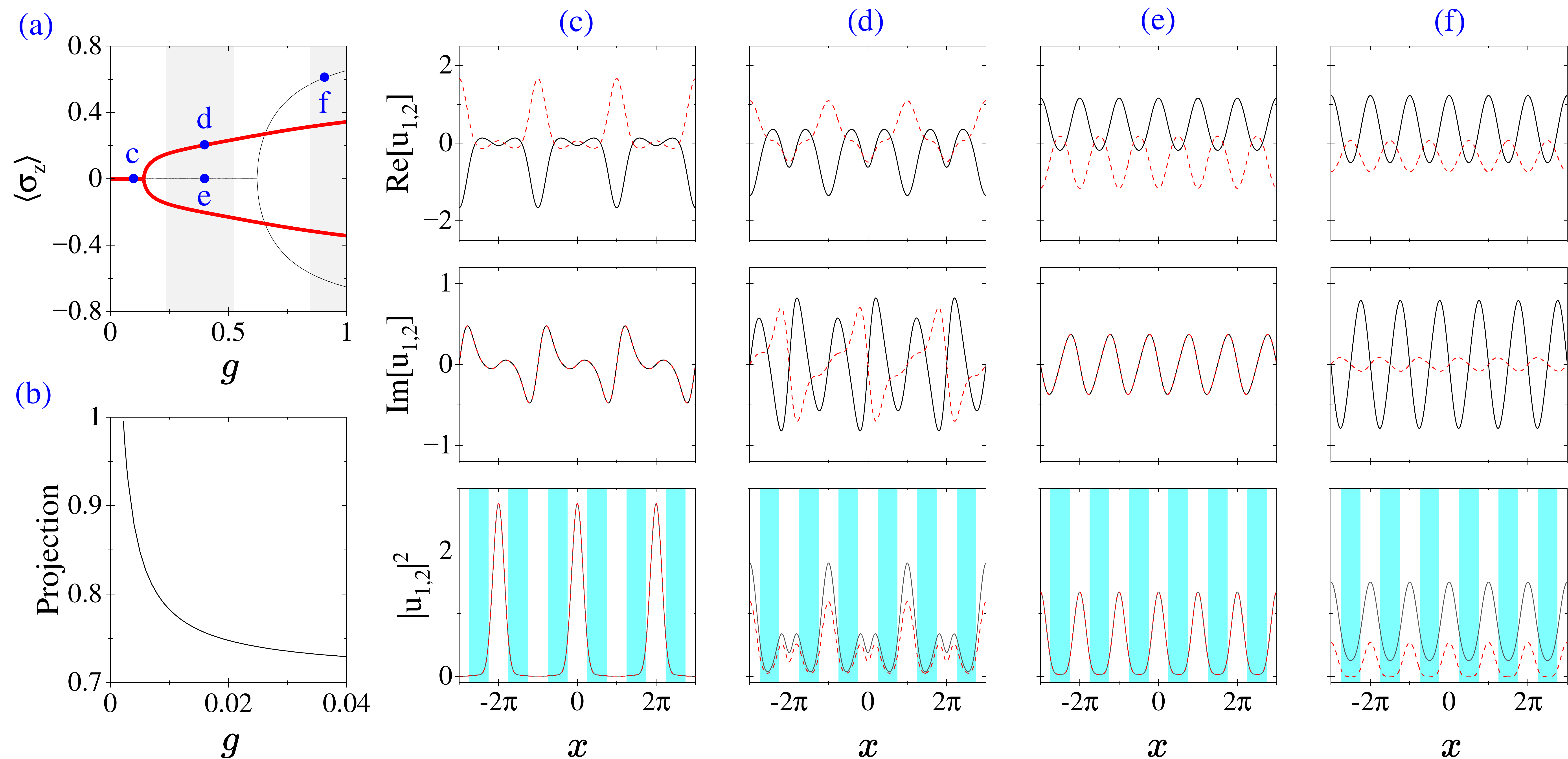}
\caption{ Existence and spontaneous symmetry breaking of the PDDWs and NBWs. (a) The spin polarization  $\langle\sigma_{z}\rangle$ of the PDDWs (the red-thick line) and NBWs (the black-thin line) shows a bifurcation relevant to the spontaneous symmetry breaking. The states in the shadow regions are dynamically stable, and other states are dynamically unstable. The profiles of states represented by labeled points are shown in (c)-(f). (c) and (d) are for the PDDWs, and (e) and (f) correspond to the NBWs. The upper and middle panels show the real and imaginary parts of the wave function, respectively. The lower panel describes the density distributions $|\bu(x)|^2$. The black solid lines are for $u_1$, and the red-dash lines are for $u_2$. The blue shadow areas represent the regions of  $\sin^{2}\left(x\right)>1/2$.
(b) The projection of the PDDWs to the associated linear Bloch wave. There is a bound value $g_b\sim3\times10^{-3}$. When $g<g_b$, the PDDWs can not exist.  In all plots, dimensionless parameters are $V_{0}=2,\gamma = 1.95$, and $\Omega = 5$.}
\label{fig_one}
\end{figure*}

Our system is based on the experimentally realizable SOC-BECs in OLs in the Ref.~\cite{Hamner2015}. Two hyperfine states of the elongated $^{87}$Rb BEC are coupled by two oppositely propagating Raman lasers with the wave number $k_{R}$. The two-photon coupling transfers the recoil momentum $2\hbar k_{R}$ to the BEC, generating the SOC, $p_x\sigma_z$,  with the SOC strength $\hbar k_{R}/m$,  $m$ the mass of atom, $p_x$ the momentum of atoms along the propagation direction of lasers, and $\sigma_z$ the standard Pauli matrix~\cite{Zheng2013,YP2016}. The implemented SOC-BEC is loaded into an OL by adiabatically ramping up the lattice lasers with the wave number $k_{L}$ along the same direction as the Raman lasers. Such the SOC-BEC in the OLs  can be described by the dimensionless Gross-Pitaevskii equation (GPE) for the spinor $\bPsi$,
\begin{align}
	\label{GP}
i\partial_t\bPsi  = \rev{H_{\text{SOC}}} \bPsi+g\left(\bPsi^\dagger\bPsi\right)\bPsi,
\end{align}
where
\begin{align}
    H_{\text{SOC}}=-\frac{1}{2}\partial_x^{2}-i\gamma\partial_x\sigma_{z}+\frac{\Omega}{2}\sigma_{x}+V_0\sin^{2}\left(x\right),
\end{align}
is the single-particle spin-orbit-coupled Hamiltonian, $\gamma=k_{R}/k_{L}$ is the SOC strength, $\Omega$ is the Rabi frequency, and $V_{0}$ is the amplitude of the OL~\cite{Hamner2015}. The units of energy and length are chosen as $2E_L$ and $1/k_L$, respectively, where $E_L=\hbar^{2}k_{L}^{2}/2m$ is the recoil energy of the lattice lasers. In the GPE, the last term describes the contact interactions. In $^{87}$Rb BEC experiments, the inter- and intra-component interactions are almost same~\cite{Hamner2015}. Therefore, we have assumed the repulsive interactions having the same coefficient, $g=2Nm\omega_{r}a_{0}/\hbar\pi k_{L}$, with $N$ being the average number of atoms inside a unit cell of OLs, $a_{0}$ being spin-independent scattering length and $\omega_{r}$ being the trap frequency along the transverse directions. 
The dimensionless order parameter $\bPsi$ satisfies the  normalization condition, 
\begin{equation}
    \frac{1}{\mathcal{S}\pi} \int_{0}^{\mathcal{S}\pi}\bPsi^\dagger\bPsi dx=1,
    \label{norm}
\end{equation}
with $\mathcal{S}=1$ for the NBWs and $\mathcal{S}=2$ for the PDDWs. Since the size of a unit cell for these two solutions is $\mathcal{S}\pi$, the normalization is performed over a unit cell. 

In the detailed calculations, we consider the typical experimental parameters~\cite{Hamner2015}: $k_{R} = 5.63\mu \text{m}^{-1} $ and $ k_{L}=2.88\mu \text{m}^{-1}$. Consequently, the dimensionless SOC strength becomes $\gamma=k_{R}/k_{L}=1.95$. The Rabi frequency $\Omega$ and the amplitude of the OL $V_0$ are free parameters that can be tuned by changing the intensity of the Raman and lattice beams, respectively. We set $V_0=2$ in the following concrete calculations.

The periodic solutions are stationary~\cite{Maluckov2012,Yang2018}, 
\begin{align}
	\label{solution}
\bPsi\left(x,t\right) =e^{-i\mu t}\bu\left(x\right),
\end{align}
where $\mu$ is the chemical potential, and $\bu(x)=\left[u_{1}(x),u_{2}(x)\right]^{T}$ is a periodic function satisfying $\bu(x+\mathcal{S}\pi)=\bu(x)$. For $\mathcal{S}=2$, the solutions become the PDDWs. For a comparison, we also study the corresponding NBW solutions in the case of $\mathcal{S}=1$. It is noticed that the corresponding NBWs have a zero quasimomentum.  The periodic functions obey the stationary GPE,
\begin{align}
	\label{GP_mu}
\mu \bu  =H_{\text{SOC}}\bu+g\left(\left|u_{1}\right|^{2}+\left|u_{2}\right|^{2}\right)\bu.
\end{align}
They are expanded on a plane-wave basis considering their periodicity,
\begin{align}
\bu& =\sum_{n=-L}^{L}\left(\begin{array}{c}
	a_{n}\\
	b_{n}
\end{array}\right)e^{i\frac{2}{\mathcal{S}}nx},
\end{align}
where $L$ is the cut-off of plane-wave modes. The plane-wave coefficients $a_n$ and $b_n$ satisfy the normalization condition $\sum_{n=-L}^{L}(\left|a_{n}\right|^{2}+\left|b_{n}\right|^{2})=1$ according to the Eq.~(\ref{norm}). We numerically solve the stationary GPE together with the normalization condition by the standard Newton relaxation method to obtain 
$a_{n},b_{n}$ and $\mu$ for the PDDWs and NBWs. Once the solutions are known, we calculate their spin polarization $\left<\sigma_{z}\right>$, which is defined as 
\begin{align}
\left<\sigma_{z}\right>=\left< \bu|\sigma_{z}|\bu\right>&=\frac{1}{\mathcal{S}\pi }\int_0^{\mathcal{S}\pi} (|u_1|^2-|u_2|^2)dx\notag \\
&= \sum_{n=-L}^{L}(\left|a_{n}\right|^{2}-\left|b_{n}\right|^{2}).
\end{align}

The dynamical stability of {these states} is an important issue and is relevant to their experimental realizations~\cite{Wu2001,Fallani2004}.  We study the stability by the linear stability analysis and assume that the solutions are perturbed by small perturbations, 
\begin{equation}
\delta u_{j}=U_{j}\left(x\right)\exp\left(iqx-i\omega t\right)+V_{j}^{*}\left(x\right)\exp\left(-iqx+i\omega^{*}t\right),
\label{pert}
\end{equation}
where $j=1,2$, $U_{j}$ and $V_{j}$ represent perturbation amplitudes, and $q$ and $\omega$ are the quasimomentum and energy of perturbations, respectively.  Substituting the perturbed solutions $e^{-i\mu t}[\bu(x) +\delta \bu(x,t) ]$ with $\delta \bu(x,t)=[ \delta u_1,\delta u_2 ]^T$ into Eq.~(\ref{GP}) and keeping linear terms with respect to the perturbation amplitudes, we obtain the following Bogoliubov-de Gennes (BdG) equation,
\begin{equation}
{\mathcal{H}_{\text{BdG}}}\bvarphi=\omega\bvarphi.
\end{equation}
Here, $\bvarphi=\left(U_{1},U_{2},V_{1},V_{2}\right)^T$, and the BdG Hamiltonian,
\begin{align}
\label{BdG}
    \mathcal{H}_{\text{BdG}}= \left(\begin{array}{cc}
    \mathcal{A}(q) & \mathcal{B}\\
    -\mathcal{B}^* & -\mathcal{A}^*(-q)\end{array}\right),
\end{align}
with
\begin{align}
    \mathcal{A}(q)&= \left(\begin{array}{cc}
    \mathcal{L}_{1}\left(q\right) & gu_{1}u_{2}^{*}+\Omega/2\\
    gu_{1}^{*}u_{2}+\Omega/2 & \mathcal{L}_{2}\left(q\right)\end{array}\right),\notag\\ 
    \mathcal{B}&=g\left(\begin{array}{cc}
    u_{1}^{2} & u_{1}u_{2}\\
    u_{1}u_{2} & u_{2}^{2}\end{array}\right), \notag
\end{align}
and 
\begin{align}  \mathcal{L}_{j}\left(q\right)=&-\frac{1}{2}\left(iq+\partial_x\right)^{2}+(-1)^{j}i\gamma\left(iq+\partial_x\right)-\mu \notag \\
&+V_{0}\sin^{2}\left(x\right)+g\left(2\left|u_{j}\right|^{2}+\left|u_{3-j}\right|^{2}\right). \notag
\end{align}
The outstanding feature {is that the BdG Hamiltonian is non-Hermitian}, which allows the existence of imaginary excitation modes in $\omega$. In the presence of imaginary modes in $\omega$, the perturbations shall grow exponentially as indicated from Eq.~(\ref{pert}). Such dynamical growth of the perturbations shall destroy the corresponding BEC states, generating dynamical instability. For the {periodic solutions}, the BdG Hamiltonian is also periodic with the period of $\mathcal{S}\pi$. We employ the plane-wave basis for $\bvarphi$, i.e., $\bvarphi=\sum_{n=-L}^{L} \bvarphi_n e^{ i2nx/\mathcal{S}}$, with the superposition coefficients $\bvarphi_n$. In the plane-wave basis, the BdG Hamiltonian can be directly diagonalized, and the excitation spectrum $\omega$ can be obtained. {The dynamical instability of the periodic solutions is checked by analyzing the existence of imaginary modes in $\omega$.}

\section{Spontaneous symmetry breaking of PDDW}
\label{state}

Applying the theoretical frame described in the previous section, we search for the PDDWs ($\mathcal{S}=2$) and the relevant NBWs ($\mathcal{S}=1$) in the lowest band. Their existence is demonstrated in Fig.\ref{fig_one}.  
For $g=0.1$, profiles of the PDDW are shown in Fig.\ref{fig_one}(c). The real and imaginary parts of $u_j$ are demonstrated in the upper and middle panels, respectively. From these two panels, we know that the real part satisfies $\text{Re}[u_1(x)]=-\text{Re}[u_2(x)]$ and has an even parity, and the imaginary part satisfies $\text{Im}[u_1(x)]=\text{Im}[u_2(x)]$ and has an odd parity. The single-particle spin-orbit-coupled Hamiltonian $H_{\text{SOC}}$ has a group of symmetries  $\{1,\hat{\alpha}_1,\hat{\alpha}_2,\hat{\alpha}_3\}$, where 
\begin{equation}
\hat{\alpha}_1=\mathcal{P}\mathcal{K}, \  \hat{\alpha}_2=\mathcal{P}\sigma_x,\ \hat{\alpha}_3=\mathcal{K}\sigma_x,
\end{equation}
with $\mathcal{P}$ being the parity operator and $\mathcal{K}$  the complex conjugate operator, i.e., $[\hat{\alpha}_j, H_{\text{SOC}}]=0$. The {state} in Fig.\ref{fig_one}(c) obeys all these symmetries and is their eigenstate. The eigenvalue equations are $ \hat{\alpha}_1 \bu(x)= \bu(x)$ with the eigen value $+1$ and $\hat{\alpha}_3 \bu(x)= -\bu(x)$ with the eigen value $-1$. From these eigenvalue equations, we know that the real part of $\bu(x)$ has an even parity and the imaginary part has an odd parity, and the relationship $\text{Re}[u_1(x)]=-\text{Re}[u_2(x)]$ and $\text{Im}[u_1(x)]=\text{Im}[u_2(x)]$ is natural.  The density distributions $|\bu(x)|^2$ are shown in the lower panel in Fig.\ref{fig_one}(c). The spin-flip symmetries $\hat{\alpha}_2$ and $\hat{\alpha}_3$ give rise to $|u_1(x)|^2=|u_2(x)|^2$. The densities have the vanishing occupation of neighbor sites of the OL, which is the signature of the PDDWs.  Due to $|u_1(x)|^2=|u_2(x)|^2$, it is easy to identify $\left<\sigma_{z}\right>=0$, which is demonstrated in Fig.\ref{fig_one}(a).

We emphasize that the family of the PDDW shown in Fig.\ref{fig_one}(c) bifurcates from the associated linear Bloch wave. We first find the associated linear Bloch wave  (which is the Bloch wave with the zero quasimomentum in the lowest band) in the absence of nonlinearity $g=0$. Then, we calculate the projection, which is defined as the overlap between the associated linear Bloch wave and the PDDW. The projection is demonstrated in Fig.\ref{fig_one}(b) as a function of the nonlinear coefficient $g$.  As $g$ decreases towards $0$, the projection goes to $1$, indicating that the family indeed bifurcates from the linear Bloch wave. There is a bound value for the coefficient, $g_b\sim3\times10^{-3}$. If $g<g_b$, the PDDW can not exist. This confirms that {they} are a pure nonlinear phenomenon and do not have a linear analog.  {It is interesting to note that the value $g_b$ is much smaller than that in the system without the SOC~\cite{Yang2018}.}

When the interaction coefficient $g$ is small, the {states} follow all symmetries of the spin-orbit coupling, the most important consequence of which is their zero spin polarization $\left<\sigma_{z}\right>=0$. We find that a large interaction coefficient can make the {states} to spontaneously break the spin-flip symmetries $\hat{\alpha}_2$ and $\hat{\alpha}_3$. Fig.\ref{fig_one}(a) illustrates the spontaneous symmetry breaking [see the red-thick line]. When the coefficient $g$ is beyond a critical value $g_c$, i.e., $g>g_c$, the {states} bifurcate into two branches. The spin polarization of the two branches is not zero and has opposite signs. A typical {solution} in one of the two branches [labeled by ``d" in Fig.\ref{fig_one}(a)] is depicted in Fig.\ref{fig_one}(d). {It} still keeps the $\hat{\alpha}_1$ symmetry and satisfies $ \hat{\alpha}_1 \bu(x)= \bu(x)$ with the eigen value being $+1$. Therefore, the parity of the real (imaginary) part is even (odd) [see the upper and middle panels in Fig.\ref{fig_one}(d)]. This state does not obey the spin-flip symmetries $\hat{\alpha}_2$ and $\hat{\alpha}_3$. As a result, $|u_1(x)|^2\ne |u_2(x)|^2$. The density distributions demonstrated in the lower panel in  Fig.\ref{fig_one}(d) clearly show the mismatch of two-component densities. The interesting feature of this symmetry-breaking state is that all sites have occupations, and populations in neighbor sites are different in order to realize the PDDWs.  This state $\bu(x)$ spontaneously breaks the spin-flip symmetries, and therefore, it is not their eigenstate. Actually, $\hat{\alpha}_2\bu(x) $ (which is the same as $\hat{\alpha}_3\bu(x)$ considering $\hat{\alpha}_3=\hat{\alpha}_1\hat{\alpha}_2$) turns to be the state in the other branch shown in Fig.\ref{fig_one}(a).  The spin polarization in the other branch becomes $\left< \hat{\alpha}_2\bu(x) |\sigma_{z} |\hat{\alpha}_2\bu(x)\right>=\left< \mathcal{P}\bu(x) |\sigma_x\sigma_{z}\sigma_x |\mathcal{P}\bu(x)\right>=- \left< \bu(x) |\sigma_{z} |\bu(x)\right>$. Therefore, the two branches have the same spin polarization but with opposite signs.

 \begin{figure}[b]
\includegraphics[width=\columnwidth]{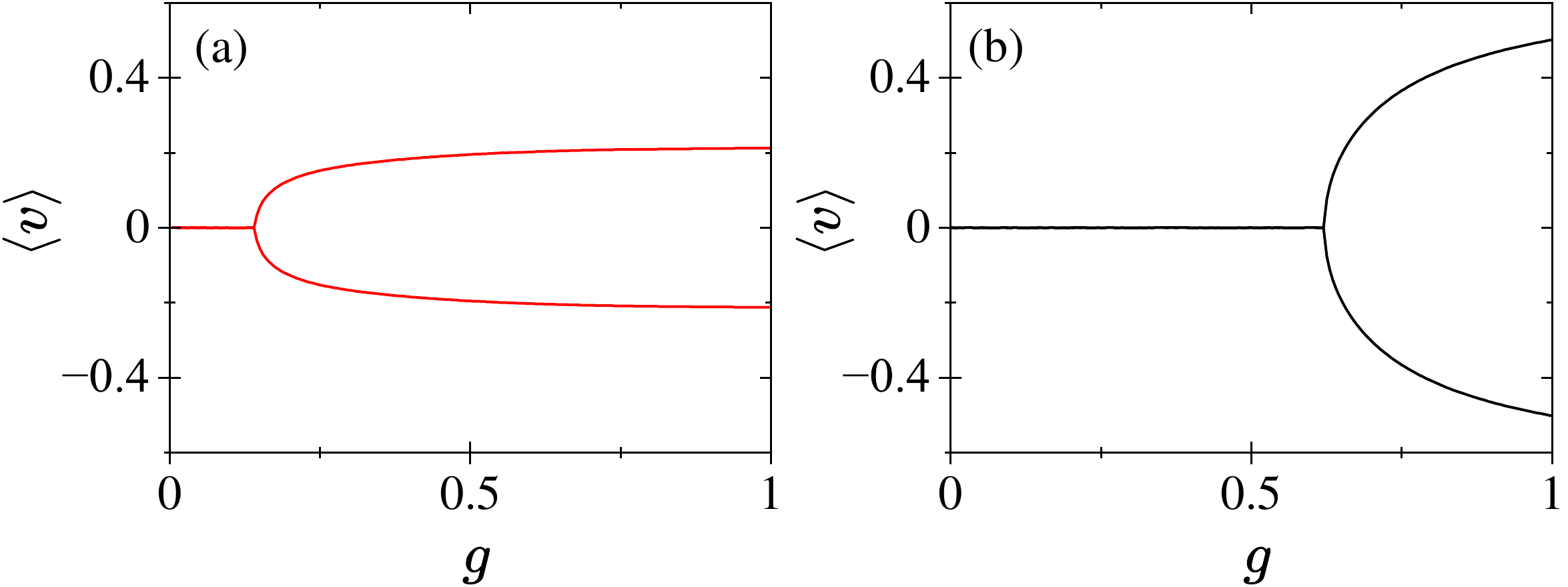}
\caption{{The average group velocity $\left<{v}\right>$ of the PDDWs (a) and the associated NBWs (b) also shows a bifurcation relevant to the spontaneous symmetry breaking. Other parameters are  $\gamma=1.95, \Omega=5$ and $V_0=2$.}}
	\label{fig_velocity}
\end{figure}

We find that the relevant NBWs have a similar behavior of spontaneous symmetry breaking as the PDDWs. The existence of the NBWs is shown by the black-thin line in 
Fig.\ref{fig_one}(a). When $g$ is not large, the NBWs follow the spin-flip symmetries and feature $\left<\sigma_{z}\right>=0$. A typical profile  is shown in 
Fig.\ref{fig_one}(e). The symmetries $\hat{\alpha}_i$ make the real (imaginary) part of the wave function to be even (odd) and generate $\text{Re}[u_1(x)]=-\text{Re}[u_2(x)]$ and $\text{Im}[u_1(x)]=\text{Im}[u_2(x)]$. Different from the PDDW in Fig.\ref{fig_one}(c), the densities $|u_1(x)|^2 = |u_2(x)|^2$ distribute in every site as shown in the lower panel of Fig.\ref{fig_one}(e).  When $g$ is large, the NBWs spontaneously break the spin-flip symmetries $\hat{\alpha}_2$ and $\hat{\alpha}_3$ and feature two branches with opposite $\left<\sigma_{z}\right>$ [see Fig.\ref{fig_one}(a)]. A profile in one of the branches is demonstrated in Fig.\ref{fig_one}(f). The real and imaginary parts in the upper and middle panels show that the NBW still follows the  $\hat{\alpha}_1$ symmetry. The densities  $|u_1(x)|^2 \neq |u_2(x)|^2$ are shown in the lower panel.

{The spontaneous symmetry breaking in Fig.\ref{fig_one}(a) is the bifurcation of nonzero $\langle\sigma_{z}\rangle$. Due to the spin-momentum locking of SOC, nonzero $\langle\sigma_{z}\rangle$ leads to the presence of group velocity $\left<{v}\right>$  ~\cite{Sherman2015,Sherman2018,Mardonov2015, Lyu2024, Xu2024}, which is defined as
\begin{align}
    \left<{v}\right>&=\left< \bu|\sigma_{z}-i\partial_x|\bu\right>\notag\\
    &=\left<\sigma_{z}\right>+\sum_{n=-L}^{L}n(\left|a_{n}\right|^{2}+\left|b_{n}\right|^{2}).
\end{align}
The nonzero $\left<{v}\right>$ means that the corresponding states carry density current. The spontaneous symmetry breaking  is also shown from the bifurcation of $\left<{v}\right>$, which is demonstrated in Fig.~\ref{fig_velocity} (a) for the PDDWs and ~\ref{fig_velocity} (b) for the associated NBWs. These two plots explain that the no-current-carrying state spontaneously breaks into two states carrying opposite current. }

\section{Stabilizing PDDW}
\label{stability}

We first explain the reason that the PDDWs without the SOC are always dynamically unstable. The typical linear Hamiltonian is $H_{\text{lin}}=-\partial_x^2/2+V_0\sin^{2}\left(x\right)$, including the kinetic energy and an OL. The Hamiltonian features Bloch band-gap spectrum $E_{nk}$ with associated Bloch waves. Here, $n$ labels the $n$-th band, and $k$ is the quasimomentum. Due to the parity symmetry $\mathcal{P}$ of $H_\text{lin}$, we have $E_{nk}=E_{n(-k)}$. Therefore, $k=0$ is a high symmetric point. The important consequence is $\partial_k E_{nk} = 0$ at the high symmetric point. So at $k=0$, $E_{nk}$ must be an extreme value due to the parity symmetry. If it is a local minimum, the effective mass of the Bloch wave at $k=0$ is positive, ${m_{\text{eff}}}\propto 1/\partial^2_k E_{nk} >0$. If it is a local maximum, the effective mass of the Bloch wave at $k=0$ is negative. Now, let's include the contact interactions and consider the PDDWs. Similar to Bloch waves, the generalized PDDWs are defined as $e^{ikx} {u_{\text{pd}}}(x)$ with the periodic functions satisfying $u_{\text{pd}}(x+2\pi)=u_{\text{pd}}(x)$. They always follow the parity symmetry of ${H_{\text{lin}}}$. Then {its} nonlinear energy at $k=0$ is an extreme value. It has been shown by numerical calculations that the \rev{state} at $k=0$ always belongs to a local maximum~\cite{Machholm2004, Seaman2005,Yang2018}. Therefore, {it} always takes {a} negative effective mass. It is known that the state with the negative effective mass is dynamically unstable in the presence of the repulsive contact interactions~\cite{Qizhong2015,Wu2001, Fallani2004,Machholm2003,Zhang2013,Smerzi2002,Danshita2007,Ozawa2013}. Such the {negative-effective-mass induced} instability can be understood as the interactions that the corresponding state effectively feels are attractive. Finally, we conclude that the PDDW at $k=0$ is dynamically unstable.  We emphasize that it is the parity symmetry of $H_{\text{lin}}$ that conserves the sign of effective mass of the {states} at $k=0$ for all different parameters.

The PDDW we consider in Eq.(\ref{solution}) belongs to the {state} at $k=0$. In the presence of the SOC, the parity must be associated with the spin. Indeed, the $H_{\text{SOC}}$ respects the spin-flip parity symmetry, $\hat{\alpha}_2=\mathcal{P}\sigma_x$. The PDDWs also obey this symmetry when the interaction coefficient $g$ is not dominant. The signature of the symmetry is $\left<\sigma_{z}\right>=0$ {and $\left<{v}\right>=0$}. This symmetry makes the {state} at $k=0$ to be an extreme value state.  Similar to the case without the SOC, {it} is a local maximum state. It is straightforward to expect that the PDDWs are dynamically unstable. We systematically check their instability by the calculation of the BdG equation. The results agree with the prediction that all {of them} having the spin-flip parity symmetry are unstable. 

 \begin{figure}[t]
\includegraphics[width=\columnwidth]{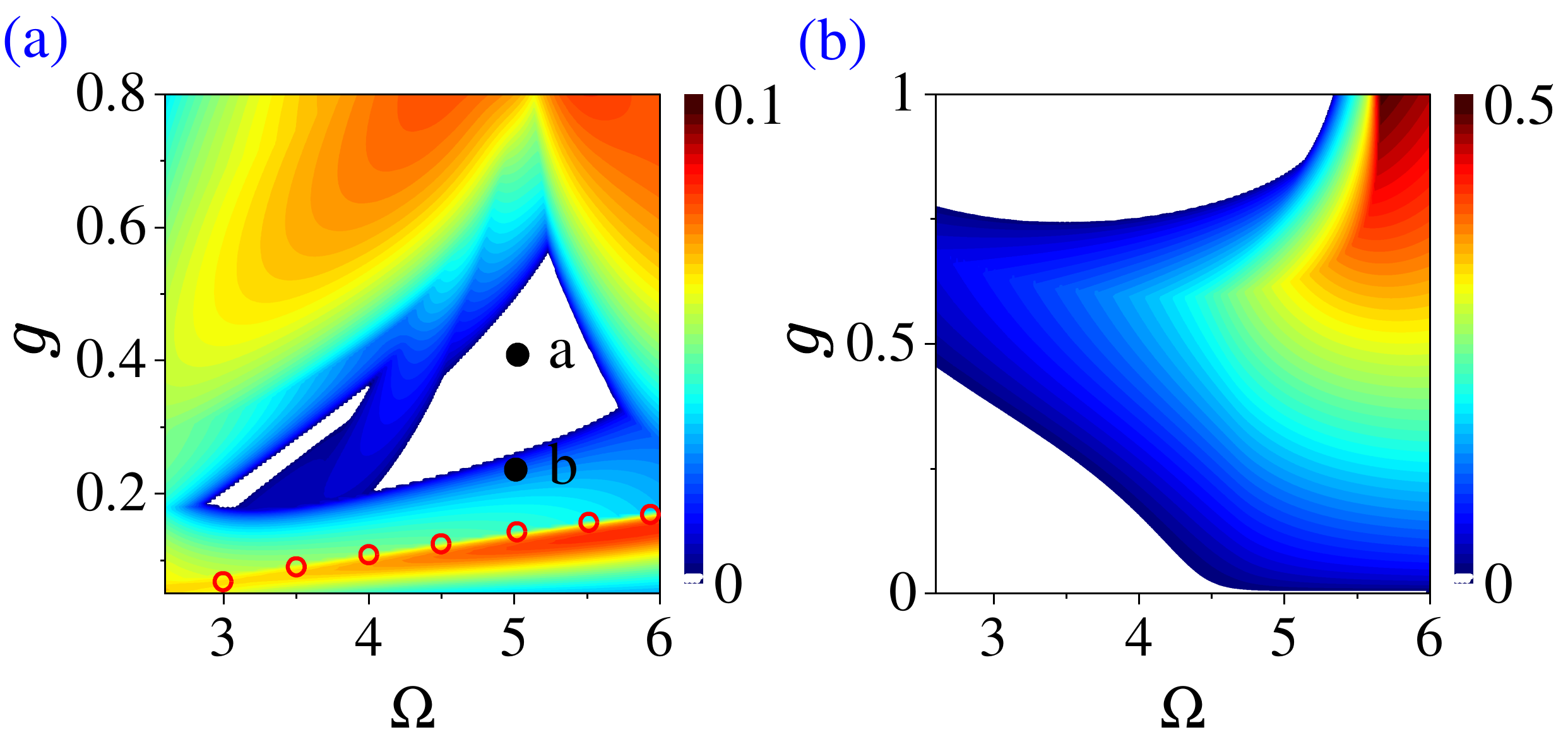}
\caption{The stability regions of the PDDWs (a) and the associated NBWs (b) in the parameter space $(\Omega,g)$. The color scale represents the maximum of the imaginary part of $\omega$, $\text{Max}(\text{Im}[\omega])$. Stable PDDWs and NBWs having $\text{Max}(\text{Im}[\omega])=0$ are in the {white}-colored regions. The open circles represent the critical values $g_c$ of the nonlinearity beyond which the PDDWs break the spin-flip parity symmetry. The two {black}-solid circles labeled by ``a" and ``b" represent the stable and unstable {states}, respectively, whose Bogoliubov spectrum and dynamical evolutions shall be demonstrated in Fig.~\ref{fig_three} and Fig.~\ref{fig_four}.   Other parameters are  $\gamma=1.95$ and $V_0=2$.}
	\label{fig_two}
\end{figure}

The spin-flip parity symmetry, $\hat{\alpha}_2=\mathcal{P}\sigma_x$, originating from the SOC, offers a possible spin channel for its breaking drown. When the interaction coefficient $g$ is dominant, the PDDWs spontaneously break the $\hat{\alpha}_2$ symmetry by not following its spin part $\sigma_x$. Because of the symmetry breaking, the {states} $\bu(x)$ and $\hat{\alpha}_2\bu(x)$ are no longer the same state. These two independent states are not the extreme value states anymore, therefore, {carry currents and} lost the guarantee that their effective mass should be negative. Once they get the positive effective mass, they might be stable considering the repulsive contact interactions. We examine the stability of the PDDWs in the two spontaneous symmetry-breaking branches shown in Fig.\ref{fig_one}(a) and do find the stable {solutions}. The stable ones are demonstrated by the red-thick lines in the shadow region in Fig.\ref{fig_one}(a).  They only exist in a finite region of $g$.  When $g$ is large beyond this region, the interactions obliterate the effect of the SOC, and the system behaves more like the one without the SOC, leading to the PDDWs becoming unstable again.  

The similar spontaneous symmetry-breaking behavior of the NBWs shown in Fig.\ref{fig_one}(a) indicates a similar stability property as the PDDWs. We check the stability of the NBWs in Fig.\ref{fig_one}(a). The NBWs having spin-flip parity symmetry are always unstable, and there is a region represented by the black-thin lines in the shadow area in the figure where the symmetry-broken NBWs are stable.

The mechanism to stabilize the PDDWs by the SOC is that the SOC makes the parity be associated with the spin-flip, and the contact interactions can break the joint symmetry by spontaneously destroying the spin-flip part to change the effective mass of the PDDWs from originally negative to positive.  The stable {states} existing in the finite region of $g$ in Fig.~\ref{fig_one}(a) stimulate us to search them in the parameter space $(\Omega, g)$.  We first find the {solutions} for different parameters $\Omega$ and $g$, and then analyze their stability by calculating the BdG equation in Eq.~(\ref{BdG}). We use the maximum of the imaginary part of $\omega$, i.e., {$\text{Max}(\text{Im}[\omega])$}, 
to show their stability. Only when $\text{Max}(\text{Im}[\omega])=0$, the corresponding {state} is stable. The results are demonstrated in Fig.~\ref{fig_two}(a), where the color scale represents the magnitude of $\text{Max}(\text{Im}[\omega])$. In the parameter space $(\Omega, g)$, there is a tilted triangle-like region in the middle of the plot (the {white}-colored region), inside which most of the PDDWs are stable. {The stable PDDWs exist in a finite region of $\Omega$. For a fixed $\Omega$, the stable region also exists in the finite region of $g$. The critical values $g_c$ for the spontaneous symmetry breaking are shown by the open circles in Fig.~\ref{fig_two}(a). It can be seen that the stable PDDWs break the spin-flip parity symmetry.} 

 \begin{figure}[t]
	\includegraphics[width=\columnwidth]{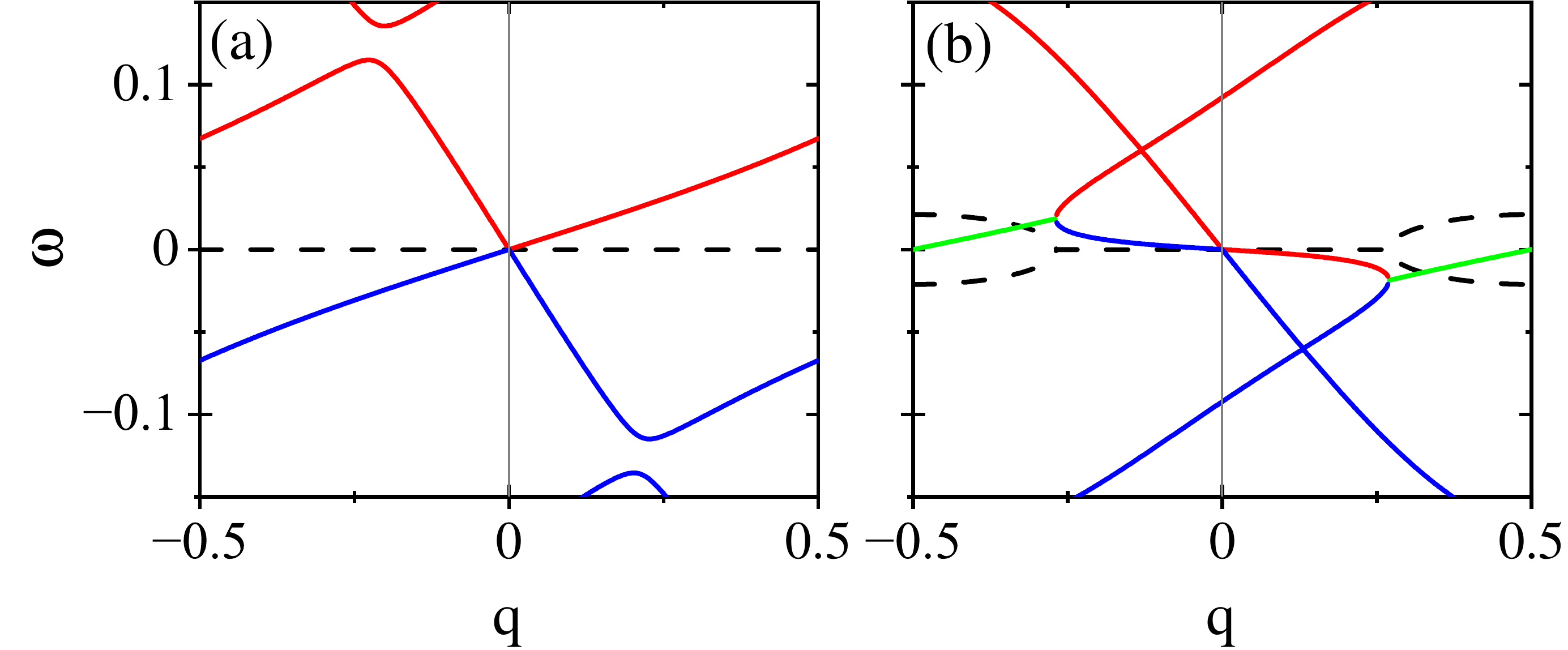}
	\caption{The Bogoliubov spectrum $\omega(q)$ of the stable (a) and unstable (b) PDDWs which are labeled by ``a" and ``b"  respectively in Fig.~\ref{fig_two}(a).  (a) $g=0.4$ and (b) $g=0.24$. The black-dashed lines are the imaginary part of $\omega$, and the red (blue) lines represent the excitations having a positive (negative) Bogoliubov norm $\mathcal{N}$. In (b), two excitations with the opposite sign of the norm collide, and after the collision, their real part merges together (the green lines), and the imaginary part appears. The grey vertical lines are for $q=0$, and the other parameters are $\gamma = 1.95$, $V_0=2$ and $\Omega = 5$. }
	\label{fig_three}
\end{figure}
\begin{figure}[b]
	\includegraphics[width=\columnwidth]{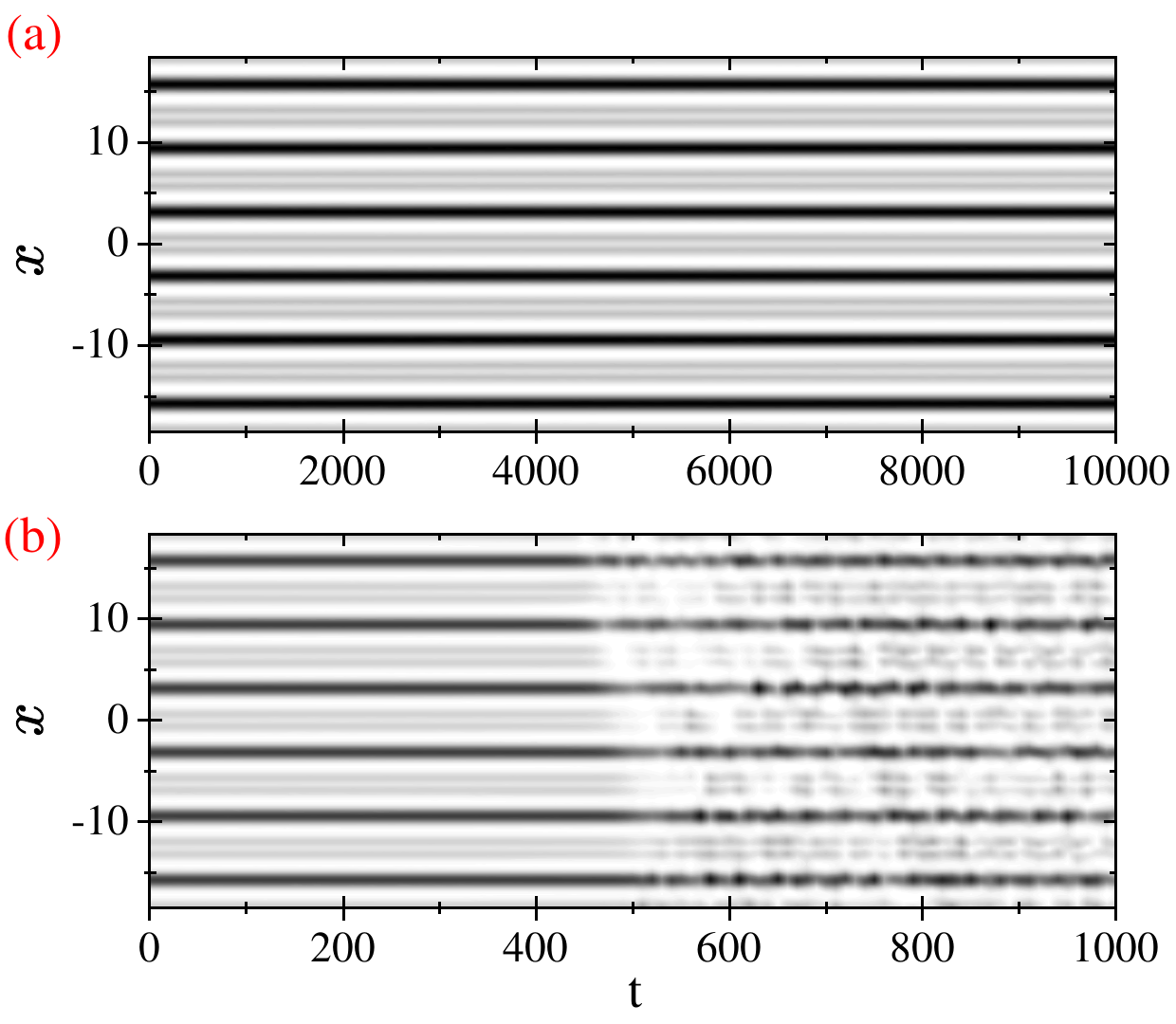}
	\caption{The nonlinear evolution of the labeled PDDWs ``a" and ``b" in Fig.~\ref{fig_two}(a) with a 5$\%$ Gaussian distributed random noise.  (a) The evolution of the stable state ``a", $g=0.4$. (b) The evolution of the unstable state ``b", $g=0.24$ and $\text{Max}(\text{Im}[\omega])=0.021$. The other parameters are $\gamma=1.95, V_{0}=2$ and $\Omega=5$. It is noticed that the time scale is different in (a) and (b). }
	\label{fig_four}
\end{figure}

A typical stable (unstable) PDDW is labeled by ``a" (``b") in Fig.~\ref{fig_two}(a).
The Bogoliubov spectrum, i.e., the Bogoliubov eigenvalue $\omega$ as a function of the quasimomentum of perturbations $q$, is demonstrated in Figs.~\ref{fig_three}(a) and~\ref{fig_three}(b) respectively for the states ``a" and ``b".  Considering the $2\pi$ period of the BdG Hamiltonian, the Bogoliubov spectrum features the Bloch band-gap structure with the first Brillouin zone $q\in (-0.5, 0.5]$. Only the lowest bands are shown in Fig.~\ref{fig_three}(a). From the BdG equation in Eq.~(\ref{BdG}), we define the Bogoliubov norm~\cite{Kawaguchi2004,Skryabin2000,Lundh2006}, 
{
\begin{align}
    \mathcal{N}&= \left< \bvarphi | \tau_z | \bvarphi\right>\notag\\
    &=  \frac{1}{2\pi}\int_0^{2\pi } dx ( |U_1|^2+|U_2|^2-|V_1|^2-|V_2|^2),
\end{align}   }
with $\tau_z=\sigma_z \otimes \mathbf{I}$, and $\mathbf{I}$ being the $2\times 2$ identity matrix.  In Fig.~\ref{fig_three}(a), the excitations in the red lines have $\mathcal{N}>0$ and in the blue lines have $\mathcal{N}<0$. The excitation with the negative norm does not have a physical consequence~\cite{Wu2001}. Therefore, we only consider the positive norm excitations in Fig.~\ref{fig_three}(a). In the lowest band and $q\rightarrow 0$, the spectrum is gapless, which is the signature of the gauge symmetry breaking. The imaginary part of $\omega$ is always zero, indicating the {state} ``a" is dynamically stable.  In contrast, for the {state} ``b", Fig.~\ref{fig_three}(b) demonstrates the presence of the imaginary part (the black-dashed lines). Two excitations with opposite signs of the norm $\mathcal{N}$ collide together, and after the collision, the real part of the two excitations merges, and the imaginary part appears. The sign of the norm is called Krein signature, and the collision of modes with opposite Krein signature can induce the generation of imaginary modes, which is the same as the phenomenon in Fig.~\ref{fig_three}(b). Such Krein collision physics is attracting research attention in the field of pseudo-Hermitian systems~\cite{Melkani,Starkov}.

The stability results shown in Fig.~\ref{fig_two}(a) can be confirmed by the nonlinear evolution to the GPE in Eq.~(\ref{GP}).  The evolution of the labeled states ``a" and ``b" in Fig.~\ref{fig_two}(a) is demonstrated in Figs.~\ref{fig_four}(a) and~\ref{fig_four}(b) respectively. For the nonlinear evolution, we incorporate a Gaussian distributed noise with the order of 5$\%$ of the initial {states}. The ``a"  state is known to be stable from the linear stability analysis. It can evolve without changing the shape for a very long time in Fig.~\ref{fig_four}(a). In contrast, the unstable {one} ``b" loses its shape during a short time evolution as shown in Fig.~\ref{fig_four}(b). {Notice that $\text{Max}(\text{Im}[\omega])$ of the state ``b" is 0.021, providing a typical time scale $t\sim300$ for happening instability.}

Finally, we also check the stability of the associated NBWs in the parameter space $(\Omega, g)$.  The results are demonstrated in Fig.~\ref{fig_two}(b) with the color scale representing the amplitude of $\text{Max}(\text{Im}[\omega])$. The stability of the NBWs has a dramatic difference with the PDDWs in Fig.~\ref{fig_two}(a).  As shown in Fig.~\ref{fig_two}(b), there are two areas in the parameter space where the NBWs are stable.

\section{Conclusion}
\label{conclusion}

In this work, we propose to use the SOC to stabilize the PDDWs in BECs with OLs.  Without the SOC, these interesting states are always unstable, the reason of which is that their negative effective mass is conserved by the parity symmetry. The SOC makes the parity to be associated with the spin flip, therefore, the joint symmetry becomes the spin-flip parity symmetry. The contact interactions can spontaneously break the spin-flip parity symmetry by destructing the spin-flip channel. The spontaneous symmetry breaking can change the effective mass of the PDDWs from originally negative to positive. With the possibly positive effective mass, the PDDWs may be dynamically stable. Instructed by this stabilization mechanism, we systematically study the spontaneous symmetry breaking of the PDDWs and check their stability in the full parameter space by analyzing the BdG equations and nonlinear evolutions. We do find the stable PDDWs existing in the experimentally accessible parameter region.

\section{ACKNOWLEDGMENTS}
This work was supported by the National Natural Science Foundation of China (NSFC) with Grants No. 112374247 and No. 11974235, and by the Shanghai Municipal Science and Technology Major Project (Grant No. 2019SHZDZX01-ZX04).

%


\end{document}